\documentclass{elsart}
\usepackage{epsfig,feynmp}

\def\beq{\begin{eqnarray}}
\def\eeq{\end{eqnarray}}
\newcommand{\lsim}{\raisebox{-4pt}{$\,\stackrel{\textstyle <}{\sim}\,$}}
\def\boldsymbol#1{{\bf #1}}

\begin{document}
\begin{frontmatter}

\begin{flushright}
WU B 97-22 \\
hep-ph/9709203 
\end{flushright}

\title{A perturbative approach \\ to the $\eta_c\gamma$ 
     transition form factor}

\author{Thorsten Feldmann\thanksref{DFG}}
and \author{Peter Kroll}

\address{Department of Theoretical Physics, University of Wuppertal,\\
    D-42097 Wuppertal, Germany}

\thanks[DFG]{Supported by {\it Deutsche Forschungsgemeinschaft}\\
 E-mail: {\tt feldmann@theorie.physik.uni-wuppertal.de}}

\begin{abstract}
The $\eta_c\gamma$ transition form factor is calculated
within a perturbative approach.
For the $\eta_c$-meson, a wave function of the
Bauer-Stech-Wirbel type is used where
the two free parameters, namely the decay constant
$f_{\eta_c}$ and the transverse size of the wave function,
are related to the Fock state probability and the
width for the two-photon decay $\Gamma[\eta_c \to \gamma\gamma]$.
The $Q^2$ dependence of the $\eta_c\gamma$ transition
form factor is well determined.
\end{abstract}

\begin{keyword}
eta/c (2980), transition form factor, hard scattering
\PACS{14.40.Gx, 12.38.Bx, 13.40.Gp}
\end{keyword}

\end{frontmatter}

\section{Introduction}
\begin{fmffile}{letterpic}
\unitlength0.8cm

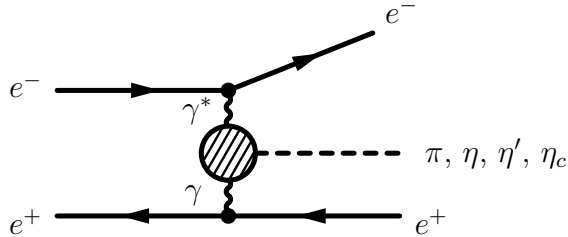
\begin{figure}[hbt]
\begin{center}
\vspace{2em}
\begin{fmfgraph*}(6,4)
\fmfpen{thick}
\fmfleft{bb,ep1,vv,em1,cc} \fmfright{aa,ep2,etac,xx,em2}
\fmf{fermion}{ep2,v2,ep1}
\fmf{fermion}{em1,v1}
\fmf{phantom}{v1,xx}
\fmffreeze
\fmf{fermion}{v1,em2}
\fmf{photon,label=$\gamma^*$}{v1,v}
\fmf{photon,label=$\gamma$  }{v,v2}
\fmffreeze
\fmf{dashes}{v,etac}
\fmfdot{v1,v2}
\fmfvn{label=$e^+$}{ep}{2} 
\fmfvn{label=$e^-$}{em}{2}
\fmfv{label=$\pi$,, $\eta$,, $\eta'$,, $\eta_c$}{etac} 
\fmfv{dec.sh=circle,dec.filled=.3,dec.si=.15w}{v}      
\end{fmfgraph*}
\end{center}
\caption{Meson-Photon transition form factors in $e^+e^-$ collisions.}
\label{Profig}
\end{figure}

In 1995 the CLEO collaboration has presented their
preliminary data on pseudoscalar meson-photon
transition form factors (see Fig.~\ref{Profig}) at large momentum transfer
$Q^2$ for the first time \cite{CLEO95}. Since then
these form factors attracted the interest of many
theoreticians and it can be said that
the CLEO measurement has strongly stimulated
the field of hard exclusive reactions. One of
the exciting aspects of the $\pi\gamma$ form factor
is that it possesses a well-established
asymptotic behavior \cite{BrLe80,WaZe73}, namely
$F_{\pi\gamma} \to \sqrt2f_\pi/Q^2$ where $f_\pi(=131~{\rm MeV})$
is the decay constant of the pion.
At the upper end of the measured $Q^2$ range 
the CLEO data \cite{CLEO95,CLEO97} only deviate by about
15\% from that limiting value. Many theoretical papers are
devoted to the explanation of that little difference.
The perhaps most important result of these analyses, as far
as they are based upon perturbative approaches
(see e.g.\ \cite{KrRa96,JaKrRa96,MuRa97}), is the rather precise
determination of the pion's light-cone wave function.
It turns out that the pion's distribution amplitude,
i.e.\ its wave function integrated over transverse
momentum, is close to the asymptotic form\footnote{Allowing 
for a second term in
the expansion of the pion's distribution amplitude upon the
the eigenfunctions of the evolution kernel in order to
quantify possible deviations from the asymptotic form,
we find, from the recent CLEO data \cite{CLEO97} and within
the modified hard scattering approach, a value of
$0.0\pm0.1$ for the expansion coefficient $B_2$ at the
scale $\mu=1$~GeV.}
($\sim x(1-x)$).
This result has far-reaching consequences for the explanation
of many hard exclusive reactions in which pions participate
(see, for instance, \cite{BrJiPaRo97,BoKrSc97a,BoKrSc97b}).

The situation is more complicated for the other cases,
the $\eta\gamma$ and the $\eta'\gamma$ form factors.
One has to determine not only the corresponding wave functions
but as well the decay constants and the $SU(3)_F$ octet-singlet
mixing angle for pseudoscalars. 
With the help of a few plausible assumptions
a determination of these quantities from the $\eta\gamma$
and $\eta'\gamma$ transition form factors seems possible
\cite{JaKrRa96}.

There is a fourth form factor of the same type, namely
the $\eta_c\gamma$ form factor which is neither
experimentally nor theoretically known. Since a
measurement of that form factor up to a momentum
transfer of about 10~GeV$^2$ seems feasible
\cite{Au96}, a theoretical analysis and prediction of it
is desirable. The purpose of this paper is the
presentation of such an analysis.
In analogy to the $\pi\gamma$ case \cite{KrRa96,JaKrRa96}
we will employ a perturbative approach on the basis
of a factorization of short- and long-distance
physics \cite{BrLe80}. 
Observables are then described as convolutions
of a so-called hard scattering amplitude to be calculated
from perturbative QCD and universal (process-independent)
hadronic light-cone wave functions, which embody soft non-perturbative
physics. The wave functions are not calculable with sufficient
degree of accuracy at present and one generally has to rely on
more or less well motivated model assumptions.

In the case of interest the mass of the charm quarks,
the $\eta_c$ meson is composed of, already
provides a large scale which allows
the application of the perturbative approach
 even for zero virtuality of the
probing photon, $Q^2\to 0$,
and, therefore, our analysis can be linked to the two-photon
decay width $\Gamma[\eta_c \to
\gamma\gamma]$. The experimental information on the
latter width provides a constraint on the $\eta_c$
wave function.
The valence Fock state probability $P_{c\bar c}$ of the $\eta_c$,
which is expected to lie in the range $0.8-1.0$, offers a
second constraint on the wave function and,
for the simple ansatz we will use, determines it completely.
We will show that variation of $P_{c\bar c}$ over the expected range has
only a very mild influence on the final result,
and hence our prediction for 
the transition form factor as a function of $Q^2$
turns out to be practically model-independent in the region of
experimental interest, where potential $Q^2$ dependence
from higher order QCD corrections can be neglected.

The organization of this paper is as follows:
First we discuss the perturbative approach to the
$\eta_c\gamma$ transition form factor, including the
leading order result for the hard scattering amplitude
and our ansatz for the wave function (sect.~\ref{sec:2}). 
In the following
sect.~\ref{sec:3} the two parameters that enter our wave function
are fixed by relating them to the Fock state probability and the
width for the two-photon decay. 
We present our results
and conclusions in sect.~\ref{sec:4}.

\section{The perturbative approach}
\label{sec:2}
In analogy to the case of the $\pi\gamma$ case
\cite{KrRa96,JaKrRa96} we define the $\eta_c\gamma$
transition form factor as a convolution of a 
hard scattering amplitude $T_H$ and a non-perturbative (light-cone)
wave function $\Psi$ of the $\eta_c$'s
leading $c\bar c$ Fock state
\beq
F_{\eta_c\gamma} (Q^2)
        &=&
\int_0^1 dx \, \int \frac{d^2 {\vec k_\perp}}{16 \pi^3} \,
        \Psi(x, {\vec k_\perp}) \, T_H(x,{\vec k_\perp},Q) \ .
\label{Fdef2}
\eeq
Here $\vec k_\perp$ denotes the transverse momentum of the
$c$ quark defined with respect to the meson's momentum and
$x$ is the usual momentum fraction carried by the $c$ quark.
In contrast to the $\pi\gamma$ case we do not include
a Sudakov factor in eq.~(\ref{Fdef2}) and therefore we are
not forced to work in the transverse configuration space.
The Sudakov factor which comprises higher order QCD
corrections in next-to-leading-log approximation
\cite{bot:89,LiSt92},
can be ignored for two reasons:
First, due to the large  mass of the $c$ quark the QCD
corrections only produce soft divergences but no collinear ones,
and hence, the characteristic double logs do not appear.
Secondly, the Sudakov factor is only relevant in the endpoint
regions ($x \to 0$ or $1$) where it provides strong
suppressions of the perturbative contribution. Since, however,
the $\eta_c$ wave function is expected to be
strongly peaked at $x=x_0$, with
$x_0=1/2$, and exponentially damped for $x \to 0,1$ the endpoint
regions are unimportant anyway.

\begin{figure}[hbt]
\begin{center}
\vspace{1em}
  \parbox{4cm}{\begin{fmfgraph}(4,2.4)
  \fmfpen{thick} \fmfleft{q1,q2} \fmfright{c,cbar}
  \fmf{photon}{q1,v1}
  \fmf{photon}{q2,v2}
  \fmf{fermion}{v1,c} 
  \fmf{fermion}{cbar,v2}
  \fmf{fermion}{v2,v1}
  \fmfdotn{v}{2}
  \end{fmfgraph}}
\hskip3em
  \parbox{4cm}{\begin{fmfgraph}(4,2.4)
  \fmfpen{thick} \fmfleft{q2,q1} \fmfright{c,cbar}
   \fmf{phantom}{q2,v1} \fmf{phantom}{q1,v2}
  \fmf{photon,tension=0}{q1,v1}
  \fmf{photon,tension=0}{q2,v2}
  \fmf{fermion}{v1,c} 
  \fmf{fermion}{cbar,v2}
  \fmf{fermion}{v2,v1}
  \fmfdotn{v}{2}
  \end{fmfgraph}}
\vspace{1em}
\end{center}
\caption{The leading order Feynman diagrams contributing
to the hard scattering amplitude for the $\eta_c\gamma$
transition form factor.}
\label{fig:feynm}
\end{figure}
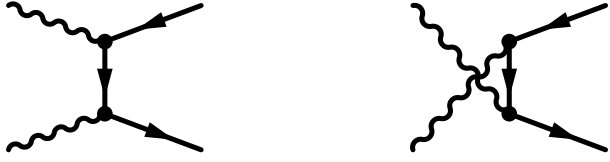
The hard scattering amplitude in leading order is 
calculated from
the Feynman diagrams shown in Fig.~\ref{fig:feynm}. 
With one photon being almost on-shell $q_1^2 \simeq 0$ and
the virtuality of the second photon denoted as $q_2^2 = -Q^2$, this
leads to (with $\bar x = (1-x)$)
\beq
T_H(x,{\vec k_\perp},Q) &=& 
\frac{e_c^2 \, 2 \sqrt 6}{x\, Q^2 + ( x \bar x + \rho^2) \, M_{\eta_c}^2 +
  {\vec k_\perp^2}}
+ (x \leftrightarrow \bar x)
+ O(\alpha_s)
\label{THhat2}
\eeq
where $M_{\eta_c}$ ($=2.98~{\rm GeV}$) is the mass of the $\eta_c$ 
meson.
$\rho$ is the ratio of the charm quark mass
($m_c$) and the $\eta_c$ mass, for which we will
take the value $\rho = 0.5$. The charge of the charm quark
in units of the elementary charge is denoted by $e_c$.
Due to the symmetry of the
wave function $\Psi(x)=\Psi(\bar x)$, 
the two graphs provide identical contributions. 

For the $\eta_c$ 
wave function we use a form adapted from
Bauer, Stech and Wirbel~\cite{BSW85},
\beq
\Psi(x, {\vec k_\perp}) = \frac{f_{\eta_c}}{2 \sqrt{6}} \,
\phi(x) \, \Sigma({\vec k_\perp}) \ .
\label{psidef}
\eeq
$f_{\eta_c}$ is the decay constant (corresponding to
$f_\pi = 131$~MeV) which plays the role of the
configuration space wave function at the origin. $\phi(x)$ is
the quark distribution amplitude which is parameterized as
\beq
 \phi(x) &=& N_\phi(a) \ x \, \bar x \
 \exp\left[ - a^2 \, M_{\eta_c}^2 \, \left(x - x_0\right)^2\right] \ .
\label{da}
\eeq
The normalization constant $N_\phi(a)$ is determined from
the usual requirement $\int_0^1 dx \, \phi(x)~=~1$.
The distribution amplitude (\ref{da}) exhibits a pronounced
maximum at $x_0$ and is exponentially damped in the endpoint
regions. This feature of the distribution amplitude parallels
the theoretically expected and experimentally confirmed
behavior of heavy hadron fragmentation functions.
Furthermore, $\Sigma$
is a Gaussian shape function which takes into
account the finite transverse size of the meson,
\beq
\Sigma({\vec k_\perp}) 
  &=& 16 \pi^2 \, a^2 \, \exp[-a^2 \, {\vec k_\perp^2}]
\ , \qquad \int \frac{d^2{\vec k_\perp}}{16\pi^3} \, \Sigma({\vec
   k_\perp}) = 1 \ .
\label{Sigmafunc}
\eeq
Frequently used and for light mesons even mandatory \cite{BHL,MuRa97} is a 
form of the $\vec k_\perp$ dependence like
$\exp[-b^2 k_\perp^2/x\bar x]$. Due to the
behavior of the distribution amplitude (\ref{da}) any
explicit appearance of $x$ in $\Sigma$ can be replaced by $x_0$
to good approximation.

\section{Fixing the parameters}
\label{sec:3}

Let us start with the determination of the $\eta_c$
decay constant, a parameter which is not accessible 
in a model-independent way at present.
Usually, one estimates $f_{\eta_c}$ through a
non-relativistic approach which provides a connection
between $f_{\eta_c}$ and the well-determined decay
constant of the $J/\psi$. We note, that the
non-relativistic approach, which is only valid
for $Q^2\ll M_{\eta_c}^2$, is consistent with our
definition of $f_{\eta_c}$ (see (\ref{Fdef2}-\ref{psidef}))
if relativistic corrections are ignored.
In the non-relativistic approach
the partial widths $\Gamma[\eta_c\to\gamma\gamma]$ 
and $\Gamma[J/\psi \to e^+e^-]$ 
are related to each other
\beq
\Gamma[\eta_c \to \gamma\gamma] &=&
    \frac{3\, e_c^4 \, \alpha^2}{m_c^2} \, |R_S(0)|^2 \
    \left[ 1 - 3.4 \, \frac{\alpha_s}{\pi}\right] \,
    \left[ 1 - \lambda_2 \, v^2 \right]^2 \ ,\nonumber \\
\Gamma[J/\psi \to e^+e^-]  &=&
    \frac{e_c^2 \, \alpha^2}{m_c^2} \, |R_S(0)|^2 \ 
    \left[1 - 5.3 \, \frac{\alpha_s}{\pi} \right] \, 
    \left[1 - \lambda_1 \, v^2 \right]^2 \ .
\label{non-rel}
\eeq
Here $R_S(r)$ is the common non-relativistic 
$S$-wave function of the $J/\psi$ and $\eta_c$ meson, 
$\lambda_{1,2}$ parameterize the leading
relativistic corrections, and the $\alpha_s$ corrections
have been calculated in \cite{Ba79}.
The wave function at the origin $R_S(0)$ is related to
the decay constants,  and in the limit $v^2 \to 0$,
$\alpha_s \to 0$ one has $f_{\eta_c} = f_{J/\psi}
= \sqrt{3/2m_c\pi} \, |R_S(0)|$ and $\Gamma[\eta_c \to \gamma\gamma]/
\Gamma[J/\psi\to e^+e^-]\simeq 3e_c^2$.
The latter decay constant is model-independently determined
from the $J/\psi$ leptonic decay width
\beq
\Gamma[J/\psi \to e^+e^-] 
 &=&
    \frac{4 \, \pi \, e_c^2\, \alpha^2 \, f_{J/\psi}^2}{3 \, M_{J/\psi}}
    = 5.26 \pm 0.37~\mbox{keV \cite{PDG96}} 
\label{eq:jpsiwidth}
\eeq
which leads to $f_{J/\psi}=409$~MeV.
However, the  $\alpha_s$ corrections in (\ref{non-rel}) are
large (depending on the value of $\alpha_s$ one prefers), 
and the relativistic corrections 
are usually
large and model-dependent (e.g.\ Chao et al.\
\cite{Hu96} find $\lambda_1 = 5/12$, $\lambda_2 = 11/12$
from a Bethe-Salpeter model). 
Estimates of the corrections typically lead to
\cite{Hu96,AhMe95,HwKi97}
 $f_{\eta_c}/f_{J/\psi}
=1.2\pm0.1$ and $\Gamma[\eta_c\to \gamma\gamma] = (5-7)$~keV.

The parameters entering
the wave function are further 
constrained by
the Fock state probability
\beq
 1 \geq P_{c\bar c} &=& 
 \int \frac{dx \, d^2{\vec k_\perp}}{16 \pi^3} \,
 \left| \Psi(x,{\vec k_\perp}) \right|^2 
%\nonumber \\[0.2em]
%&\simeq & 
\simeq
\frac{f_{\eta_c}^2  a^2 \, \pi^2}{3} \cdot 
{{aM_{\eta_c}} \over {\sqrt{2\,\pi }}} 
%\cdot
\left(1+\frac{2}{a^2M_{\eta_c}^2}\right)
%{{{a^4}\,{M_{\eta_c}^4}   - 2\,{a^2}\,{M_{\eta_c}^2} +  3}\over 
%    {{{ ({a^2}\,{M_{\eta_c}^2} -2)   }^2}}} 
\ .
\eeq
As we said in the introduction, 
one expects $0.8 \leq P_{c\bar c} < 1$ for a charmonium
state (for smaller values of $P_{c\bar c}$ one would not
understand the
success of non-relativistic potential models for these states). 
Since the perturbative contribution to the $\eta_c\gamma$
form factor only
mildly depends on the value of $P_{c\bar c}$, as it will
turn out below,
we use $P_{c\bar c}=0.8$ as a constraint for the transverse size
parameter $a$.
For $f_{\eta_c}=409$~MeV this leads 
to $a = 0.97$~GeV$^{-1}$,
a value that is consistent with estimates  for the radius
$\langle r^2 \rangle = 3 \,a^2 \simeq (0.4~{\rm fm})^2$ or the 
quark velocity $v^2 =3/(Ma)^2 \simeq 0.3$ from potential
models \cite{BuTy81}.

The 
two photon decay width
$\Gamma[\eta_c\to\gamma\gamma]$, the experimental value
of which still suffers
from large uncertainties~\cite{PDG96},
can be directly related to the $\eta_c\gamma$ transition 
form factor at $Q^2=0$
\beq
&& \Gamma[\eta_c \to \gamma\gamma] =
\frac{\pi \alpha^2 M_{\eta_c}^3}{4} 
   \left| F_{\eta_c\gamma}(0) \right|^2  
=\left\{\begin{array}{l}
 7.5^{+1.6}_{-1.4}~\mbox{keV} \quad \mbox{(direct)} \\  
(4.0 \pm 1.5~\mbox{keV}) \!\cdot\!  \frac{\Gamma^{\rm
    tot}_{\eta_c}}{13.2~{\rm MeV}} \quad \mbox{(BR)}
\end{array}  \right. 
\label{eq:etacwidth}
\eeq
One may use this decay rate as a normalization
condition for $F_{\eta_c\gamma}(Q^2=0)$
and present the result in the
form $F_{\eta_c\gamma}(Q^2)/F_{\eta_c\gamma}(0)$.
In this way the perturbative QCD corrections at $Q^2=0$ to
the $\eta_c\gamma$ transition form factor,
which are known to be large (see eq.~(\ref{non-rel})), are automatically
included, and also the uncertainties
in the present knowledge of $f_{\eta_c}$ do not enter
our predictions.

\section{Results and Conclusions}
\label{sec:4}
\begin{figure}[hbtp]
\unitlength1cm
\begin{center}
\hskip-.5cm
{\psfig{file=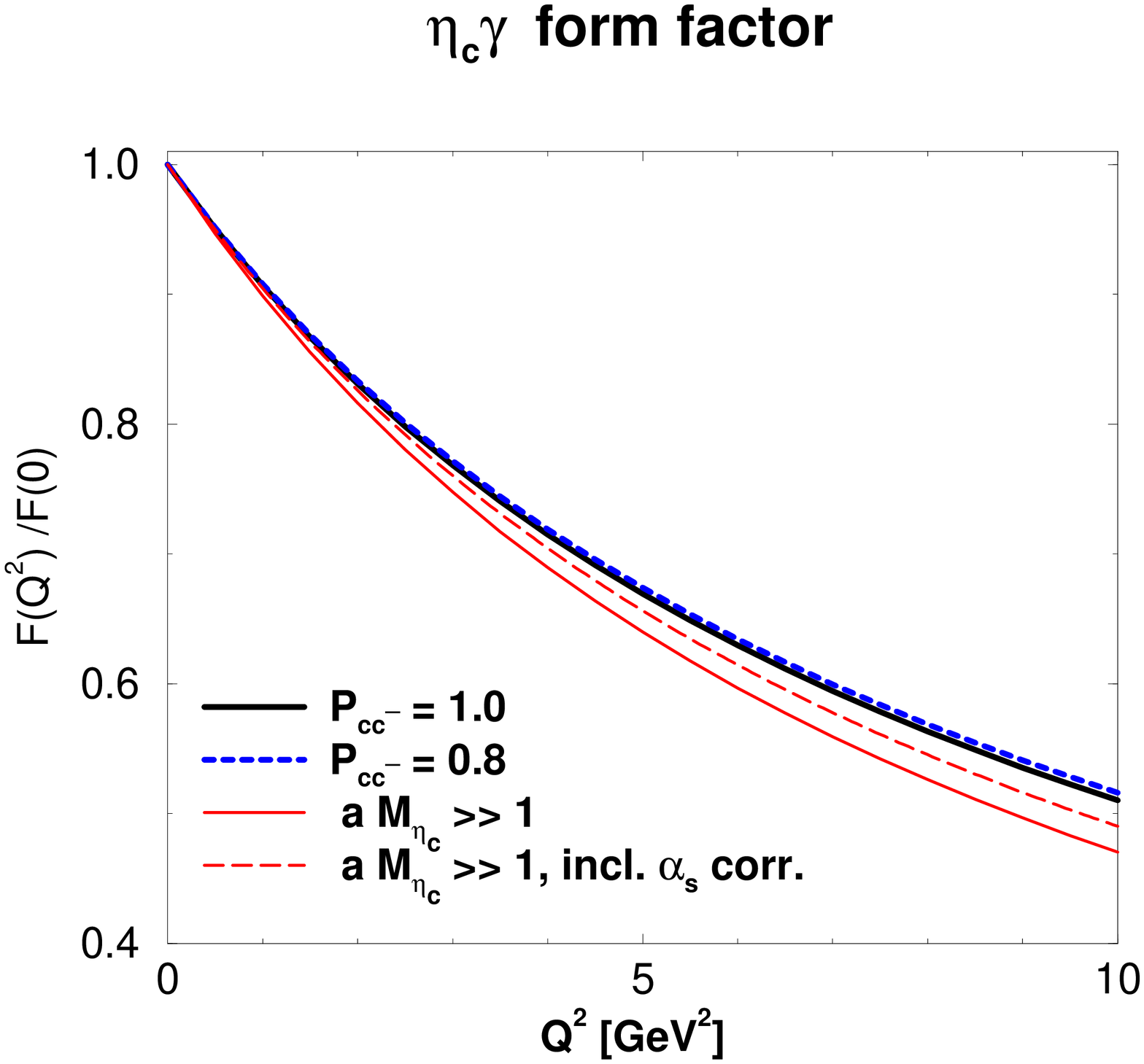, width=6.1cm, angle = -90}}\hskip-.5cm
{\psfig{file=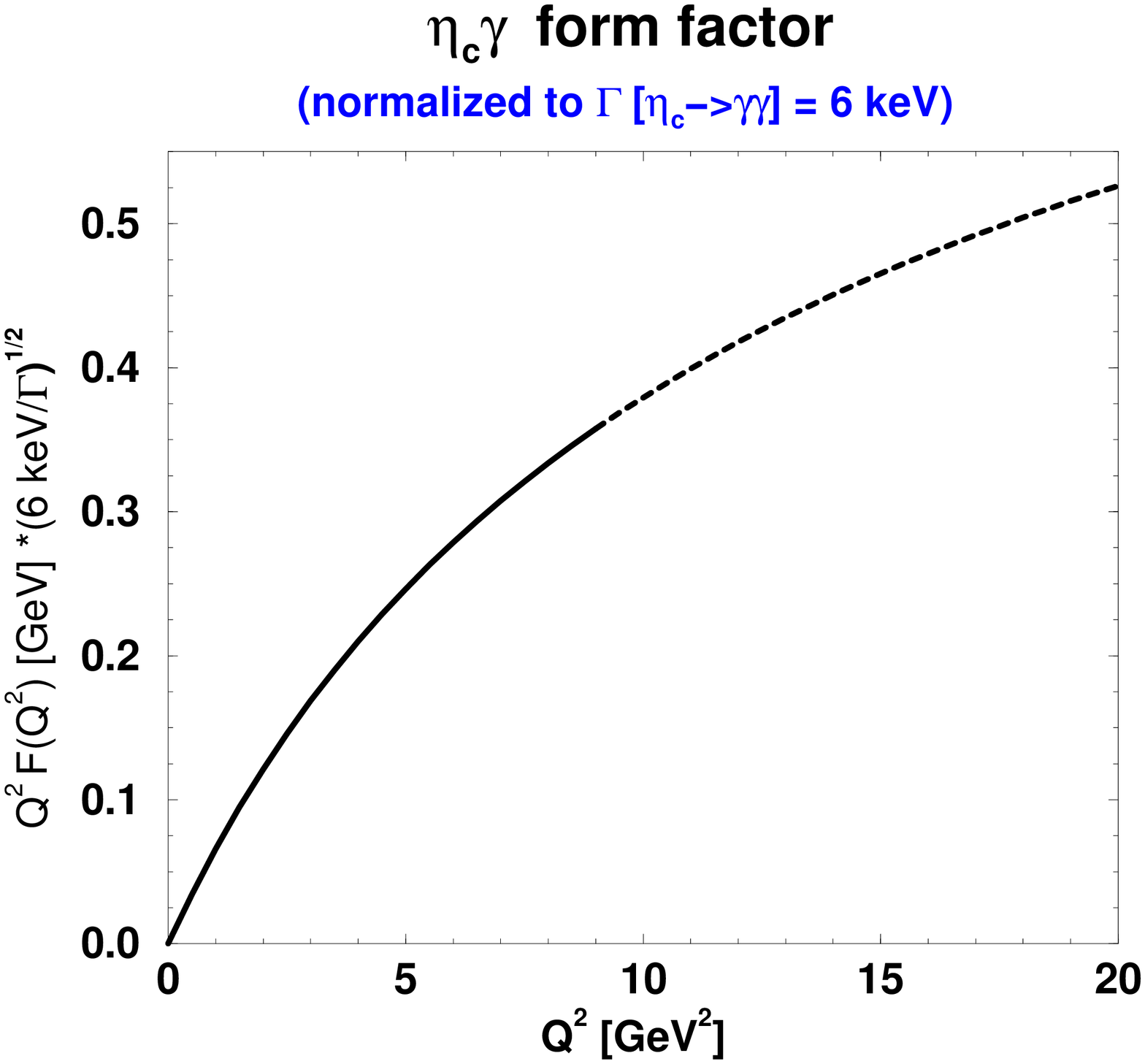, width=6.1cm, angle = -90}}
\end{center}
\caption{Left: The results $F_{\eta_c\gamma}(Q^2)/F_{\eta_c\gamma}(0)$ 
  for different values of the Fock state probability
 $P_{c\bar c}$ as well as the
approximation (\ref{compact2}) (with
$\langle \vec k_\perp^2 \rangle \to 0$) with and without
$\alpha_s$ corrections.
Right: The predictions for  
 $Q^2 \, F_{\eta_c\gamma}(Q^2)$  scaled to
 $\Gamma[\eta_c\to\gamma\gamma] = 6$~keV in the leading order of the 
perturbative approach (for $P_{q\bar q}$=0.8).
The dashes indicate the $Q^2$ region where QCD corrections may
alter the predictions slightly.}
\label{fig}
\end{figure}

Let us turn now to numerical estimates of the
$\eta_c\gamma$ transition form factor.
The left hand side of Fig.~\ref{fig} shows that form
factor for two different values of the Fock state probability 
$P_{c\bar c}$. 
As already mentioned, we observe that the dependence
on $P_{c\bar c}$ is weak. 
On the right hand side of Fig.~\ref{fig} we present the result for the
transition form factor $Q^2 \, F_{\eta_c\gamma}$
scaled to a partial width
$\Gamma[\eta_c\to\gamma\gamma]$ of $6$~keV. 

For the values of the meson mass and the transverse size parameter $a$
that we are dealing with (i.e.\ $(aM_{\eta_c})^2=9$)
it makes also sense to 
consider the peaking approximation in which the hard
scattering amplitude is evaluated at the position of the
maximum of the distribution amplitude. The peaking approximation
is formally equivalent to the replacement of the distribution
amplitude (\ref{da}) by a $\delta$ function at $x=x_0$
(into which it collapses in the limit $(aM_{\eta_c}) \to \infty$).
This approximation is numerically  quite reliable 
and allows one to discuss the qualitative
features of the model results in a rather simple fashion.
By means of the uncertainty principle 
$(aM_{\eta_c})^2 \gg 1$ can be turned
into 
$\langle k_\perp^2\rangle \ll M_{\eta_c^2}$, and
hence one may also neglect the $k_\perp^2$ dependence in
the hard scattering amplitude (collinear approximation).
Including $1/aM_{\eta_c}$ corrections to both, 
the peaking approximation and the collinear
approximation, and using  for the mean transverse
momentum the relation $1/a^2=2\langle \vec k_\perp^2 \rangle$
(see (\ref{Sigmafunc})) one arrives at the approximate
result for $Q^2\lsim M_{\eta_c}^2$
\beq
 F_{\eta_c\gamma}(Q^2) &\simeq&
       \frac{4\,e_c^2 \, f_{\eta_c} }
{Q^2 + M_{\eta_c}^2 + 2 \, \langle \vec k_\perp^2 \rangle 
}
%\ , 
%\label{compact}
%\eeq
%
%Eq.~(\ref{compact}) which may also be cast into the form
%\beq
%F_{\eta_c\gamma}(Q^2) 
\simeq
      \frac{F_{\eta_c\gamma}(0)}{1 + Q^2/(M_{\eta_c}^2 + 2\, \langle
        \vec k_\perp^2 \rangle)}
%\ , \qquad (Q^2\lsim M_{\eta_c}^2)
\ .
\label{compact2}
\eeq
which agrees with the perturbative result to order
$(1/aM_{\eta_c})^2$.
Eq.~(\ref{compact2})
reveals that, to a very good approximation, the predictions
for the $\eta_c\gamma$ form factor are rather insensitive to
the details of the wave function. Only the mean transverse
momentum following from it is required. 

To assess the quality of the approximation (\ref{compact2})
we compare it for the special case of $\langle k_\perp^2\rangle=0$
to the full result from the perturbative approach in  Fig.~\ref{fig}
(left hand side). We observe that, with increasing $Q^2$, the
two results growingly deviate from each other, at 
$Q^2=10$~GeV$^2$ the difference amounts to 10\%.
If one uses our estimated value of $\langle k_\perp^2\rangle$
in (\ref{compact2}) the deviation from the full result is
further reduced and amounts only to 4\%  at 
$Q^2=10$~GeV$^2$. This little difference is likely smaller
than the expected experimental errors in a future 
measurement of the $\eta_c\gamma$ form factor (see \cite{Au96}).
These considerations nicely illustrate
that the $Q^2$ dependence of the $\eta_c\gamma$ form factor
is well determined.
The main uncertainty
of the prediction resides in the normalization, i.e.\ the
$\eta_c$ decay constant or the value of the form factor at
$Q^2=0$.

Eq.~(\ref{compact2}) resembles the Brodsky-Lepage interpolation
formula proposed for the $\pi\gamma$ transition form factor
\cite{BLint} as well as the prediction from the vector
meson dominance model (VDM). Our value of 
$\sqrt{M_{\eta_c}^2 + 2 \langle k_\perp^2\rangle}$ is
$3.15~{\rm GeV}$ which is very close to the  value of
the $J/\psi$ mass that one would have inserted in the
VDM ansatz\footnote{Considering the uncertainty in
the mean transverse momentum and in the $c$-quark mass
(see (\ref{THhat2})), we estimate the uncertainty in
the effective pole position following from
(\ref{compact2}) to amount to about 5\%.}.
In the VDM the $\eta_c\gamma$ form factor at
$Q^2=0$ is given by $F_{\eta_c\gamma}^{\rm VDM}(0)=
e_c \, g_{J/\psi\eta_c\gamma} \, f_{J/\psi}/M_{J/\psi}$
where the $J/\psi\eta_c\gamma$ coupling constant can be
obtained from the radiative decay $J/\psi \to \eta_c\gamma$
\cite{PDG96,Du83}. One finds $F_{\eta_c\gamma}^{\rm VDM}(0)=
0.048$~MeV$^{-1}$ and hence $\Gamma^{VDM}[\eta_c\to
\gamma\gamma]=2.87$~keV which appears to be somewhat
small as compared to the experimental values quoted
in (\ref{eq:etacwidth}). Inclusion of  a similar 
contribution form the $\psi'$ pole does not improve the
VDM result  since
the $\psi'$ contribution is very small. 
In the case of two virtual photons $q_1^2\neq 0$
the perturbative
 prediction $F(q_1^2,q_2^2)\propto 1/(-q_1^2-q_2^2 + M_{\eta_c}^2 +
2 \langle \vec k_\perp^2 \rangle)$ differs substantially from the VDM.

Let us briefly discuss, how $\alpha_s$ corrections 
may modify the leading order result for the
$\eta_c\gamma$ form factor:
One has to consider two distinct kinematic regions.
First, if $Q^2 \lsim M_{\eta_c}^2$ one can neglect the
evolution of the wave function, and one is left with the
QCD corrections to the hard scattering amplitude $T_H$,
which have been calculated in the peaking and collinear
approximation to order $\alpha_s$ in \cite{ShVy81}.
For the scaled form factor the $\alpha_s$ corrections
at $Q^2$ and at $Q^2=0$ cancel to a high degree, and
even at $Q^2=10$~GeV$^2$ the effect of the $\alpha_s$
corrections is less than 5\% (see left side of
Fig.~\ref{fig}).

Secondly, for $Q^2 \gg M_{\eta_c}^2$, one can neglect the quark and
meson masses and arrives at the same situation as for 
the pions. The $\alpha_s$ corrections to the hard 
scattering amplitude and the evolution of the wave function
with $Q^2$ are known \cite{BrLe80,ShVy81,Br83}.
For very large values
of $Q^2$ the asymptotic behavior of the transition form factor
is completely determined by QCD, since any meson distribution
amplitude evolves into the asymptotic form
$\phi(x)\to\phi_{\rm as}(x)=6 \, x \, \bar x$, 
\beq
F_{\eta_c\gamma}(Q^2) \to 
\frac{2 e_c^2 f_{\eta_c}}{Q^2} \,\int_0^1 dx \, \frac{\phi(x)}{x}
\to \frac{8 \, f_{\eta_c}}{3 \, Q^2} \ , \qquad
(Q^2 \to \infty)
\label{asymp}
\eeq
The value of the moment $\langle x^{-1}\rangle=\int dx\,\phi(x)/x$
evolves from $2.5$ 
to the asymptotic
value $3$. We note that the asymptotic behavior of the peaking
approximation 
(\ref{compact2}) 
is $16/9 f_{\eta_c}/Q^2$. The deviation
from (\ref{asymp}) demonstrates the inaccuracy of 
the peaking approximation for broad distribution amplitudes.

A precise measurement of the strength of the $\eta_c\gamma$
transition form factor may serve to determine the decay constant
$f_{\eta_c}$ (see, e.g.\ 
(\ref{compact2})
). Though attention
must be paid to the fact that the obtained value of $f_{\eta_c}$
is subject to large QCD corrections (about of the order
10-15\% for $Q^2\lsim 10$~GeV$^2$) which should be taken into
account for an accurate extraction of the $\eta_c$ decay constant.

\section*{Acknowledgments}
 We thank M.\ Raulfs for his contributions
in the early stage of this work and
M.\ Kienzle for drawing our attention to the $\eta_c\gamma$
problem. T.F.\ acknowledges fruitful discussions with T.\ Ohl and
 V.\ Savinov.

\bibliography{ref,kroll}

\begin{thebibliography}{10}

\bibitem{CLEO95}
CLEO collaboration, V.~Savinov {\em et~al.},
\newblock (1995), hep-ex/9507005; in proceedings of the PHOTON97 workshop,
  Sheffield (1995), eds.\ D.J.\ Miller et al., World Scientific.

\bibitem{BrLe80}
G.~P. Lepage and S.~J. Brodsky,
\newblock Phys. Rev. {\bf D22}, 2157 (1980).

\bibitem{WaZe73}
T.~F. Walsh and P.~Zerwas,
\newblock Nucl. Phys. {\bf B41}, 551 (1972).

\bibitem{CLEO97}
CLEO collaboration, V.~Savinov {\em et~al.},
\newblock (1997), CLNS preprint 97/1477.

\bibitem{KrRa96}
P.~Kroll and M.~Raulfs,
\newblock Phys. Lett. {\bf B387}, 848 (1996).

\bibitem{JaKrRa96}
R.~Jakob, P.~Kroll, and M.~Raulfs,
\newblock J. Phys. {\bf G22}, 45 (1996).

\bibitem{MuRa97}
I.~V. Musatov and A.~V. Radyushkin,
\newblock (1997), hep-ph/9702443.

\bibitem{BrJiPaRo97}
S.~J. Brodsky, C.-R. Ji, A.~Pang, and D.~G. Robertson,
\newblock (1997), hep-ph/9705221.

\bibitem{BoKrSc97a}
J.~Bolz, P.~Kroll, and G.~A. Schuler,
\newblock Phys. Lett. {\bf B392}, 198 (1997).

\bibitem{BoKrSc97b}
J.~Bolz, P.~Kroll, and G.~A. Schuler,
\newblock (1997), hep-ph/9704378, to be published in Z. Phys. C.

\bibitem{Au96}
P.~Aurenche {\em et~al.},
\newblock (1996), hep-ph/9601317.

\bibitem{bot:89}
J.~Botts and G.~Sterman,
\newblock Nucl. Phys. {\bf B325}, 62 (1989).

\bibitem{LiSt92}
H.~Li and G.~Sterman,
\newblock Nucl. Phys. {\bf B381}, 129 (1992).

\bibitem{BSW85}
M.~Wirbel, B.~Stech, and M.~Bauer,
\newblock Z. Phys. {\bf C29}, 637 (1985).

\bibitem{BHL}
S.~J. Brodsky, T.~Huang, and G.~P. Lepage,
\newblock In Banff 1981, Proceedings, Particles and Fields 2, 143- 199.

\bibitem{Ba79}
R.~Barbieri, E.~d'Emilio, G.~Curci, and E.~Remiddi,
\newblock Nucl. Phys. {\bf B154}, 535 (1979).

\bibitem{PDG96}
Particle Data Group, R.~M. Barnett {\em et~al.},
\newblock Phys. Rev. {\bf D54}, 1 (1996).

\bibitem{Hu96}
K.-T. Chao, H.-W. Huang, J.-H. Liu, and J.~Tang,
\newblock (1996), hep-ph/9601381.

\bibitem{AhMe95}
M.~R. Ahmady and R.~R. Mendel,
\newblock Z. Phys. {\bf C65}, 263 (1995).

\bibitem{HwKi97}
D.~S. Hwang and G.-H. Kim,
\newblock (1997), hep-ph/9703364.

\bibitem{BuTy81}
W.~Buchm{\"u}ller and S.~H.~H. Tye,
\newblock Phys. Rev. {\bf D24}, 132 (1981).

\bibitem{BLint}
S.~J. Brodsky and G.~P. Lepage,
\newblock Phys. Rev. {\bf D24}, 1808 (1981).

\bibitem{Du83}
O.~Dumbrajs {\em et~al.},
\newblock Nucl. Phys. {\bf B216}, 277 (1983).

\bibitem{ShVy81}
M.~A. Shifman and M.~I. Vysotskii,
\newblock Nucl. Phys. {\bf B186}, 475 (1981).

\bibitem{Br83}
E.~Braaten,
\newblock Phys. Rev. {\bf D28}, 524 (1983).

\end{thebibliography}

\end{fmffile}
\end{document}